\documentclass[10pt,a4paper]{article}

\usepackage{amsmath,amsgen,latexsym}
\usepackage{amstext,amssymb,amsfonts,latexsym}
\usepackage{theorem}
\usepackage{pifont}
\usepackage[dvips]{graphics,epsfig}
\usepackage[dvips]{graphicx}

 \newcommand{\bs}{\bigskip}
 \newcommand{\ms}{\medskip}
 \newcommand{\n}{\noindent}
 \newcommand{\s}{\smallskip}
 \newcommand{\hs}[1]{\hspace*{ #1 mm}}
 \newcommand{\vs}[1]{\vspace*{ #1 mm}}

 \newcommand{\setempty}{\mathrm{\O}}
 
 \newcommand{\nat}{\mathbb{N}}

 \newcommand{\dom}{\mbox{dom}}
 
 \newcommand{\ran}{\mbox{ran}}

 \newcommand{\co}{\mathrm{co}\mbox{-}}

 \newcommand{\ie}{\textrm{i.e.},\hspace*{2mm}}
 \newcommand{\eg}{\textrm{e.g.},\hspace*{2mm}}
 
 \newcommand{\etalc}{\textrm{et al.}}

 \newcommand{\CC}{{\cal C}}
 \newcommand{\FF}{{\cal F}}

 \newcommand{\GG}{{\cal G}}

 \newcommand{\PP}{{\cal P}}

 \newcommand{\p}{\mathrm{P}}
 \newcommand{\np}{\mathrm{NP}}

 \newcommand{\reg}{\mathrm{REG}}
 \newcommand{\cfl}{\mathrm{CFL}}
 
 \newcommand{\dcfl}{\mathrm{DCFL}}

\theoremstyle{plain}
\theoremheaderfont{\bfseries}
 \newtheorem{theorem}{Theorem}[section]
 \newtheorem{lemma}[theorem]{Lemma}
 \newtheorem{proposition}[theorem]{Proposition}

 {\theorembodyfont{\rmfamily}
  }
 {\theorembodyfont{\rmfamily} }
 {\theorembodyfont{\rmfamily} }

 \newenvironment{proofsketch}{\par \noindent
            {\bf Proof Sketch. \hs{2}}}{\hfill$\Box$ \vspace*{3mm}}

\newcommand{\ignore}[1]{}

\newcommand{\track}[2]{[\:\begin{subarray}{c} #1 \\%
      #2 \end{subarray} ]}

\newcommand{\cent}{|\!\! \mathrm{c}}
\newcommand{\dollar}{\$}

 \newcommand{\cflsvt}{\mathrm{CFLSV_t}}
 \newcommand{\cflsv}{\mathrm{CFLSV}}
  \newcommand{\cflsvtwo}{\mathrm{CFLSV}(2)}

 \newcommand{\cflmvt}{\mathrm{CFLMV_t}}
 \newcommand{\cflmv}{\mathrm{CFLMV}}
 \newcommand{\cflmvtwo}{\mathrm{CFLMV}(2)}

 \newcommand{\cflmvk}[1]{\mathrm{CFLMV}( #1 )}
 \newcommand{\cflsvk}[1]{\mathrm{CFLSV}( #1 )}

 \newcommand{\npsv}{\mathrm{NPSV}}
 \newcommand{\npmv}{\mathrm{NPMV}}

 \newcommand{\deltacfl}[1]{\Delta^{\mathrm{CFL}}_{ #1 }}
 \newcommand{\sigmacfl}[1]{\Sigma^{\mathrm{CFL}}_{ #1 }}
 \newcommand{\picfl}[1]{\Pi^{\mathrm{CFL}}_{ #1 }}

 \newcommand{\sigmacflmv}[1]{\Sigma^{\mathrm{CFL}}_{ #1 }\mathrm{MV}}
 \newcommand{\sigmacflsv}[1]{\Sigma^{\mathrm{CFL}}_{ #1 }\mathrm{SV}}

 \newcommand{\sigmacflsvt}[1]{\Sigma^{\mathrm{CFL}}_{ #1 }\mathrm{SV_t}}
 \newcommand{\picflmv}[1]{\Pi^{\mathrm{CFL}}_{ #1 }\mathrm{MV}}

 \newcommand{\optcfl}{\mathrm{OptCFL}}
 
 \newcommand{\optnfa}{\mathrm{OptNFA}}
 \newcommand{\optp}{\mathrm{OptP}}
 \newcommand{\npsvt}{\mathrm{NPSV_t}}
 \newcommand{\nfamv}{\mathrm{NFAMV}}
 \newcommand{\nfasv}{\mathrm{NFASV}}
 \newcommand{\nfasvt}{\mathrm{NFASV_t}}

\begin{document}

\pagestyle{plain}
\setcounter{page}{1}

\begin{center}
{\Large {\bf
Structural Complexity of
Multi-Valued Partial Functions \s\\ Computed by
Nondeterministic Pushdown Automata\footnote{This extended abstract has already appeared in the Proceedings of the 15th Italian Conference of Theoretical Computer Science (ICTCS 2014), September 17--19, Perugia, Italy, CEUR Workshop Proceedings, vol.1231, pp.225--236, 2014.}}} \ms\\
{\large {\bf (Extended Abstract)}}
\bs\s\\

{\sc Tomoyuki Yamakami}\footnote{Present Affiliation: Department of Information Science, University of Fukui, 3-9-1 Bunkyo, Fukui 910-8507, Japan} \bs\\
\end{center}


\begin{abstract}
This paper continues a systematic and comprehensive study on the structural
properties of CFL functions, which are in general multi-valued partial
functions computed by one-way one-head nondeterministic pushdown automata
equipped with write-only output tapes (or pushdown transducers), where CFL
refers to a relevance to context-free languages. The CFL functions tend to
behave quite differently from their corresponding context-free languages. We
extensively discuss containments, separations, and refinements among various
classes of functions obtained from the CFL functions by applying
Boolean operations, functional composition, many-one relativization, and
Turing relativization. In particular, Turing relativization helps construct
a hierarchy over the class of CFL functions.
We also analyze the computational complexity of optimization functions,
which are to find optimal values of CFL functions, and discuss their
relationships to the associated languages.

\s

\n{\bf Keywords:}
multi-valued partial function, oracle, Boolean operation,
refinement, many-one relativization, Turing relativization, CFL hierarchy,
optimization, pushdown automaton, context-free language
\end{abstract}

\section{Much Ado about Functions}\label{sec:introduction}

In a traditional field of formal languages and automata, we have dealt
primarily with {\em languages} because of their practical applications to, for
example, a parsing analysis of programming languages. Most fundamental
languages listed in many undergraduate textbooks are, unarguably, regular (or
rational) languages and context-free (or algebraic) languages.

Opposed to the recognition of languages, translation of words, for example,
requires a mechanism of transforming input words to output words. Aho
\etalc~\cite{AHU67} studied machines that produce words on output tapes
while reading symbols on an input tape.
Mappings on strings (or word relations) that are realized by such machines are
known as transductions.
Since languages are regarded, from an integrated viewpoint, as Boolean-valued
(\ie $\{0,1\}$-valued) total functions, it seems more essential to study the
behaviors of those functions. This task is, however, quite challenging, because
these functions often
demand quite different concepts, technical tools, and proof arguments, compared
to those for languages. When underlying machines are particularly {\em
nondeterministic},
they may produce numerous distinct output values (including the case of no
output values). Mappings realized by such machines become, in general,  {\em
multi-valued partial functions} transforming each admissible string to a
certain finite (possibly empty) set of strings.

Based on a model of polynomial-time nondeterministic Turing machine,
computational complexity theory has long discussed the structural complexity
of
various NP function classes, including NPMV, NPSV, and NPSV$_\mathrm{t}$ (where
MV and SV respectively stand for ``multi-valued'' and ``single-valued'' and the
subscript ``t''  does for ``total''). See, \eg  a survey \cite{Sel96}.

Within a scope of formal languages and automata, there is also rich literature
concerning the behaviors of nondeterministic finite automata equipped with
write-only output tapes (known as rational transducers) and properties of
associated multi-valued partial functions (known also as rational
transductions). Significant efforts have been made over the years
to understand the functionality of such functions.
A well-known field of functions include ``CFL functions,'' which were formally
discussed in 1963 by Evey \cite{Eve63} and Fisher \cite{Fis63} and general
properties have been since then discussed in, e.g.,  \cite{CC83,KSY07}. CFL
functions are generally computed by {\em one-way one-head nondeterministic
pushdown automata} (or {\em npda's}, in short) {\em equipped with write-only
output tapes}. For example, the function $PAL_{sub}(w)=\{x\in\{0,1\}^*\mid
\exists u,v\,[w=uxv],x=x^R\}$ for every $w\in\{0,1\}^*$ is a CFL function,
where $x^R$ is $x$ in reverse.
As subclasses of CFL functions, three fundamental function classes $\cflmv$,
$\cflsv$, and $\cflsvt$ were recognized explicitly in \cite{Yam11} and further
explored in \cite{Yam09}.

In recent literature, fascinating structural properties of $\cfl$ functions
have been  extensively investigated. Konstantinidis \etalc~\cite{KSY07} took an
algebraic approach toward the characterization of CFL functions. In relation to
cryptography, it was shown that there exists a pseudorandom generator in
$\cflsvt$ that ``fools'' every language in  a non-uniform (or an advised)
version of  $\reg$ \cite{Yam11}.
Another function class $\cflmvtwo$ in \cite{Yam09} contains pseudorandom
generators against a non-uniform analogue of  $\cfl$.
The behaviors of functions in those function classes seem to look quite
different from what we have known for context-free languages. For instance, the
single-valued total function class $\cflsv$ can be seen as a functional
extension of the language family $\cfl\cap\co\cfl$ rather than $\cfl$
\cite{Yam09}. In stark contrast with a tidy theory of $\np$ functions,
circumstances that surround $\cfl$ functions differ significantly
because of mechanical constraints (\eg a use of stacks, one-way moves of tape
heads, and $\lambda$-moves) that harness the behaviors of underlying npda's
with output tapes. One such example is concerning a notion of {\em refinement}
\cite{Sel94} (or uniformization \cite{KSY07}).
Unlike language families, a containment between two multi-valued partial
functions is customarily replaced by refinement.
Konstantinidis et al.~\cite{KSY07} posed a basic question of whether every
function in $\cflmv$ has a refinement in $\cflsv$. This question was lately
solved negatively \cite{Yam14b} and this result clearly contrasts a situation
that a similar relationship is not known to hold between $\npmv$ and $\npsv$.

Amazingly, there still remains a vast range of structural properties that await
for further investigation. Thus, we wish to continue a coherent and systematic
study on the structural behaviors of the aforementioned CFL functions. This
paper aims at exploring fundamental relationships (such as, containment,
separation, and refinement) among the above function classes and their natural
extensions via four typical operations: (i) Boolean operations, (ii) functional
composition, (iii) many-one relativization, and (iv) Turing relativization.
The last two operations are a natural generalization of {\em many-one and
Turing CFL-reducibilities} among languages \cite{Yam14a}.
We use the Turing relativization to introduce a hierarchy of function classes
$\sigmacflmv{k}$, $\picflmv{k}$, and $\sigmacflsv{k}$ for each level $k\geq1$,
in which the first $\Sigma$-level classes coincide with $\cflmv$ and $\cflsv$,
respectively.
We show that all functions in this hierarchy have linear space-complexity. With
regard to refinement, we show that, if the CFL hierarchy of \cite{Yam14a}
collapses to the $k$th level, every function in $\sigmacflmv{k+1}$ has a
refinement in $\sigmacflsv{k+1}$ for every index $k\geq2$.

Every nondeterministic machine with an output tape can be naturally seen as a
process of generating a set of feasible ``solutions'' of each instance. Among
those solutions, it is useful to study the complexity of finding  ``optimal''
solutions. This gives rise to {\em optimization functions}. Earlier, Krentel
\cite{Kre88} discussed properties of $\optp$ that is composed of
polynomial-time optimization functions. Here, we are focused on similar
functions induced by npda's with output tapes. We denote by $\optcfl$ a class
of those optimization CFL functions. This function class is proven to be
located between $\cflsvt$ and $\sigmacflsvt{4}$. Moreover, we show the class
separation between $\cflsvt$ and $\optcfl$.

To see a role of functions during a process of recognizing languages,
K\"{o}bler and Thierauf \cite{KT94} introduced a {\em $/\!/$-advice operator}
by generalizing the Karp-Lipton $/$-advice operator, and they argued the
computational complexity of languages in $\p/\!/\FF$ and $\np/\!/\FF$ induced
by any given function class $\FF$. Likewise, we discuss the complexity of
$\reg/\!/\FF$ and $\cfl/\!/\FF$ when $\FF$ are various subclasses of $\cfl$
functions, particularly $\cflsvt$ and $\optcfl$.

All omitted or abridged proofs, because of the page limit, will appear shortly
in a complete version of this paper.

\section{A Starting Point}\label{sec:preliminaries}

\vs{-1}
\n{\bf Formal Languages.}\hs{2}
Let $\nat$ be the set of all {\em natural numbers} (\ie nonnegative integers)
and set $\nat^{+}=\nat-\{0\}$. Throughout this paper, we use the term ``
polynomial'' to mean polynomials on $\nat$ with nonnegative coefficients. In
particular, a {\em linear polynomial} is of the form $ax+b$ with $a,b\in\nat$.
The notation $A-B$ for two sets $A$ and $B$ indicates the {\em set difference}
$\{x\mid x\in A, x\not\in B\}$. Given any set $A$, $\PP(A)$ denotes the {\em
power set} of $A$.
A set $\Sigma^k$ (resp., $\Sigma^{\leq k}$), where $k\in\nat$, consists only of
strings of length $k$ (resp., at most $k$).
Here, we set $\Sigma^* = \bigcup_{k\in\nat}\Sigma^k$.
The {\em empty string} is always denoted $\lambda$.  Given any language $A$
over $\Sigma$, its {\em complement} is $\Sigma^*- A$, which is also denoted by
$\overline{A}$ as long as $\Sigma$ is clear from the context.

We adopt a {\em track notation} $\track{x}{y}$ from \cite{TYL10}. For two
symbols $\sigma$ and $\tau$, $\track{\sigma}{\tau}$ expresses a new symbol. For
two strings $x=x_1x_2\cdots x_n$ and $y=y_1y_2\cdots y_n$ of length $n$,
$\track{x}{y}$ denotes a string $\track{x_1}{y_1}\track{x_2}{y_2}\cdots
\track{x_n}{y_n}$. Whenever $|x|\neq|y|$, we follow a convention introduced in
\cite{TYL10}: if $|x|<|y|$, then $\track{x}{y}$ actually means
$\track{x\#^m}{y}$, where $m=|y|-|x|$ and $\#$ is a designated new symbol.
Similarly, when $|x|>|y|$, the notation $\track{x}{y}$ expresses
$\track{x}{y\#^m}$ with $m=|x|-|y|$.

As our basic computation model, we use {\em one-way one-head nondeterministic
pushdown automata} (or npda's, in short) allowing $\lambda$-moves (or
$\lambda$-transitions) of their input-tape heads.
The notations $\reg$ and $\cfl$ stand for the families of all regular languages
and of all context-free languages, respectively. For each number
$k\in\nat^{+}$, the {\em $k$-conjunctive closure} of $\cfl$, denoted by $\cfl(k)$,
is defined to be $\{ \bigcap_{i=1}^{k}A_i \mid A_1,A_2,\ldots,A_k\in\cfl\}$
(see, \eg \cite{Yam11}).

Given any language $A$ (used as an oracle),  $\cfl_{T}^{A}$ (or $\cfl_{T}(A)$)
expresses a collection of all languages recognized by npda's equipped with
write-only query tapes with which the npda's make non-adaptive oracle queries to $A$, provided that all computation paths of the npda's must terminate in
$O(n)$ steps no matter what oracle is used \cite{Yam14a}.  We use the notation
$\cfl_{T}^{\CC}$ (or $\cfl_{T}(\CC)$) for language family $\CC$ to denote the
union $\bigcup_{A\in\CC}\cfl_{T}^{A}$. Its deterministic version is expressed
as $\dcfl_{T}^{\CC}$.
The {\em CFL hierarchy}  $\{\deltacfl{k},\sigmacfl{k},\picfl{k}\mid
k\in\nat^{+}\}$ is composed of classes $\deltacfl{1}=\dcfl$,
$\sigmacfl{1}=\cfl$, $\picfl{k}=\co\sigmacfl{k}$, $\deltacfl{k+1} =
\dcfl_{T}(\picfl{k})$, and $\sigmacfl{k+1}=\cfl_{T}({\picfl{k}})$ for each
index $k\geq1$  \cite{Yam14a}.

\ms
\n{\bf Functions and Refinement.}\hs{2}
Our terminology associated with {\em multi-valued partial functions} closely
follows the standard terminology in computational complexity theory. Throughout
this paper, we adopt the following convention: the generic term ``function''
always refers to ``multi-valued partial function,'' provided that {\em
single-valued} functions are viewed as a special case of multi-valued functions
and, moreover, partial functions include {\em total} functions. We are
interested in multi-valued partial functions mapping{}\footnote{To describe a
multi-valued partial function $f$, the expression
``$f:\Sigma^*\rightarrow\Gamma^*$'' is customarily used in the literature.
However, the current expression ``$f:\Sigma^*\rightarrow\PP(\Gamma^*)$''
matches a fact that the outcome of $f$ on each input string in $\Sigma^*$ is a
subset of $\Gamma^*$.}
$\Sigma^*$ to $\PP(\Gamma^*)$ for certain alphabets $\Sigma$ and $\Gamma$. When
$f$ is single-valued, we often write $f(x)=y$ instead of $y\in f(x)$.
Associated with $f$, $\dom(f)$ denotes the {\em domain} of $f$, defined to be
$\{x\in\Sigma^*\mid f(x)\neq\setempty\}$. If $x\not\in\dom(x)$, then $f(x)$ is said
to be {\em undefined}. The {\em range} $\ran(f)$ of $f$ is a set
$\{y\in\Gamma^*\mid f^{-1}(y)\neq\setempty\}$.

For any language $A$, the {\em characteristic function} for $A$, denoted by
$\chi_{A}$, is a function defined as $\chi_{A}(x)=1$ if $x\in A$ and
$\chi_{A}(x)=0$ otherwise. We also use a {\em quasi-characteristic function}
$\eta_{A}$, which is defined as $\eta_{A}(x)=1$ for any string $x$ in $A$ and
$\eta_{A}(x)$ is not defined for all the other strings $x$.

Concerning all function classes discussed in this paper, it is natural to
concentrate only on functions whose outcomes are bounded in size by fixed
polynomials. More precisely, a multi-valued partial function
$f:\Sigma^*\rightarrow \PP(\Gamma^*)$ is called {\em polynomially bounded}
(resp., {\em linearly bounded}) if there exists a polynomial (resp., a linear
polynomial) $p$ such that, for any two strings $x\in\dom(f)$ and
$y\in\Gamma^*$, if $y\in f(x)$ then $|y|\leq p(|x|)$ holds. In this paper, we
understand that {\em all function classes are made of polynomially-bounded
functions}.
Given two alphabets $\Sigma$ and $\Gamma$, a function
$f:\Sigma^*\rightarrow\PP(\Gamma^*)$ is called {\em length preserving} if, for
any $x\in\Sigma^*$ and $y\in\Gamma^*$, $y\in f(x)$ implies $|x|=|y|$.

Whenever we refer to a {\em write-only tape}, we always assume that (i)
initially, all cells of the tape are blank, (ii) a tape head starts at the
so-called {\em start cell}, (iii) the tape head steps forward whenever it
writes down any non-blank symbol, and (iv) the tape head can stay still only in
a blank cell. An {\em output} (outcome or output string) along a computation
path is a string produced on the output tape after the  computation path is
terminated. We call an output string {\em valid} (or {\em legitimate}) if it is
produced along a certain accepting computation path.

\sloppy
To describe npda's, we take the following specific definition. For any given
function $f:\Sigma^*\rightarrow\PP(\Gamma^*)$, an npda $N$ equipped with a one-way
read-only input tape and  a write-only output tape that computes $f$ must have
the  form
$(Q,\Sigma,\{\cent,\dollar\},\Theta,\Gamma,\delta,q_0,Z_0,Q_{acc},Q_{rej})$,
where $Q$ is a set of inner states, $\Theta$ is a stack alphabet, $q_0$ ($\in
Q$) is the initial state, $Z_0$ ($\in\Theta$) is the stack's bottom marker, and
$\delta:(Q-Q_{halt})\times (\check{\Sigma}\cup\{\lambda\})\times \Theta
\rightarrow \PP(Q\times \Theta^*\times (\Gamma\cup\{\lambda\}))$ is a
transition function, where $Q_{halt} = Q_{acc}\cup Q_{rej}$,
$\check{\Sigma}=\Sigma\cup\{\cent,\dollar\}$, and $\cent$ and $\dollar$ are
respectively left and right endmarkers.
It is important to note that,  in accordance with
the basic setting of \cite{Yam14a},  we consider only {\em npda's whose computation
paths are all terminate (\ie reach halting states) in $O(n)$
steps},\footnote{If no execution time bound is imposed, a function computed by
an npda that nondeterministically produces every binary string on its output
tape for each input becomes a valid CFL function; however, this function is no
longer an $\np$ function.} where $n$ refers to input size. We refer to this specific condition
regarding execution time as the {\em termination condition}.

A function class $\cflmv$ is composed of all (multi-valued partial) functions
$f$, each of which maps $\Sigma^*$ to $\PP(\Gamma^*)$ for certain alphabets
$\Sigma$ and $\Gamma$ and there exists an npda $N$ with a one-way read-only
input tape and a write-only output tape such that, for every input
$x\in\Sigma^*$,  $f(x)$ is a set of all {\em valid} outcomes of $N$ on the
input $x$. The termination condition imposed on our npda's obviously
leads to an anticipated containment $\cflmv\subseteq \npmv$. Another class
$\cflsv$ is a subclass of $\cflmv$ consisting of single-valued partial
functions. In addition, $\cflmvt$ and $\cflsvt$ are respectively restrictions
of $\cflmv$ and $\cflsv$ onto {\em total functions}. Those function classes
were discussed earlier in \cite{Yam11}.

An important concept associated with multi-valued partial functions is {\em
refinement} \cite{Sel94}. This concept  (denoted by $\sqsubseteq_{ref}$) is
more suitable to use than {\em set containment} ($\subseteq$).
Given two multi-valued partial functions $f$ and $g$, we say that $f$ is a {\em
refinement} of $g$, denoted by $g\sqsubseteq_{ref}f$, if  (1) $\dom(f)=\dom(g)$
and (2) for every $x$, $f(x)\subseteq g(x)$ (as a set inclusion).
We also say that $g$ is {\em refined} by $f$. Given two sets
$\FF$ and $\GG$ of functions, $\GG\sqsubseteq_{ref} \FF$ if  every function
$g$ in $\GG$ can be refined by a certain function $f$ in $\FF$.
When $f$ is additionally single-valued, we call $f$ a {\em single-valued
refinement} of $g$.

\section{Basic Operations for Function Classes}\label{sec:function-class}

Let us discuss our theme of the structural complexity of various classes of (multi-valued partial)
functions by exploring fundamental relationships among those function classes.
In the course of our discussion, we will unearth an exquisite
nature of CFL functions, which looks quite different from that of NP functions.

We begin with demonstrating basic features of CFL functions.  First, let us present close  relationships between CFL functions and context-free languages. Notice that, for any function $f$ in $\cflmv$, both $\dom(f)$ and $\ran(f)$ belong to
$\cfl$.
It is useful to recall from \cite{Yam14a} a notion of $\natural$-extension. Assuming that $\natural\not\in\Sigma$, a {\em
$\natural$-extension} of a given string $x\in\Sigma^*$ is a string $\tilde{x}$
over $\Sigma\cup\{\natural\}$ satisfying the following requirement: $x$ is obtained directly from
$\tilde{x}$ simply by removing all occurrences of $\natural$ in $\tilde{x}$.
For example, if $x=01101$, then $\tilde{x}$ may be $01\natural 1\natural01$ or
$\natural 0110\natural\natural 1$. The next lemma resembles Nivat's
representation theorem for rational transductions (see, \eg \cite[Theorem
2]{KSY07}).

\begin{lemma}\label{translate-CFLMV-to-CFL}
For any function $f\in\cflmv$, there exist a language $A\in \cfl$ and a linear
polynomial $p$ such that, for every pair $x$ and $y$, $y\in f(x)$ iff  (i)
$\track{\tilde{x}}{\tilde{y}}\in A$, (ii) $|y|\leq p(|x|)$, and  (iii)
$|\tilde{x}|=|\tilde{y}|$ for certain strings $\tilde{x}$ and $\tilde{y}$,
which are $\natural$-extensions of $x$ and $y$, respectively.
\end{lemma}

An immediate application of Lemma \ref{translate-CFLMV-to-CFL} leads to a
functional version of the well-known pumping lemma \cite{BPS61}.

\begin{lemma}\label{func-pumping-CFLMV}{\rm (functional pumping lemma for
$\cflmv$)}
Let $\Sigma$ and $\Gamma$ be any two alphabets and let
$f:\Sigma^*\rightarrow\PP(\Gamma^*)$ be any function in $\cflmv$. There exist
three numbers $m\in\nat^{+}$ and $c,d\in\nat$ that satisfy the following
condition. Any string $w\in\Sigma^*$ with $|w|\geq m$ and any string $s\in
f(w)$ are decomposed into $w= uvxyz$ and $s=abpqr$ such that (i) $|vxy|\leq m$,
(ii) $|vybq|\geq1$, (iii) $|bq|\leq cm+d$, and (iv) $ab^ipq^ir\in f(uv^ixy^iz)$
for any number $i\in\nat$. In the case where $f$ is further length preserving, the following condition also holds: (v) $|v|=|b|$ and $|y|=|q|$. Moreover,
(i)--(ii) can be replaced by (i') $|bq|\geq1$.
\end{lemma}

Boolean operations over languages in $\cfl$ have been extensively discussed in
the past literature (e.g.,  \cite{Wot78,Yam14a}). Similarly, it is possible to
consider Boolean operations that are directly applicable to functions.
In particular, we are focused on three typical Boolean operations: union,
intersection, and complement. Let us define the first two operations.
Given two function classes $\FF_1$ and $\FF_2$, let $\FF_1\wedge \FF_2$ (resp.,
$\FF_1\vee \FF_2$) denote a class of all functions $f$ defined as
$f(x)=f_1(x)\cap f_2(x)$ (set intersection) (resp., $f(x)= f_1(x)\cup f_2(x)$,
set union) over all inputs $x$, where $f_1\in\FF_1$ and $f_2\in\FF_2$.
Expanding $\cflmv(2)$ in Section \ref{sec:introduction}, we inductively define
a {\em $k$-conjunctive function class} $\cflmvk{k}$ as follows: $\cflmv(1) =
\cflmv$ and $\cflmv(k+1)=\cflmv(k)\wedge \cflmv$ for any index $k\in\nat^{+}$.
Likewise,  $\cflsvk{k}$ is defined using $\cflsv$ instead of $\cflmv$.

\begin{proposition}\label{cflsvk+1-vs-cflsvk}
Let $k,m\geq1$.
\begin{enumerate}\vs{-1}
  \setlength{\topsep}{-2mm}%
  \setlength{\itemsep}{0mm}
  \setlength{\parskip}{0cm}%

\item $\cflmv(\max\{k,m\}) \subseteq \cflmv(k)\vee \cflmv(m) \subseteq
    \cflmv(km)$.

\item $\cflmv(\max\{k,m\}) \subseteq \cflmv(k)\wedge \cflmv(m) \subseteq
    \cflmv(k+m)$.

\item $\cflsvk{k}\subsetneqq\cflsvk{k+1}$.

\end{enumerate}
\end{proposition}

Note that Proposition \ref{cflsvk+1-vs-cflsvk}(3) follows indirectly from a result in \cite{LW73}.

Fenner \etalc~\cite{FHOS97} considered ``complements'' of $\np$ functions.
Likewise, we can discuss {\em complements of CFL functions}.
Let $\FF$ be any family of functions whose output sizes are upper-bounded by
certain {\em linear polynomials}.  A function $f$ is in $\co\FF$ if there are
a linear polynomial $p$, another function $g\in\FF$, and a constant
$n_0\in\nat$ such that, for any pair $(x,y)$ with $|x|\geq n_0$,
$y\in f(x)$ iff both $|y|\leq p(|x|)$ and $y\not\in g(x)$ holds.
This condition
implies that $f(x) = \Sigma^{\leq a|x|+b} - g(x)$ (set difference) for all
$x\in\Sigma^{\geq n_0}$. The finite portion $\Sigma^{<n_0}$ of inputs is
ignored.

The use of set difference in the above definition makes us introduce another
class operator $\ominus$. Given two sets $\FF,\GG$ of functions,
$\FF\ominus\GG$ denotes a collection of functions $h$ satisfying the following:
for certain two functions $f\in\FF$ and $g\in\GG$, $h(x) =
f(x) - g(x)$ (set difference) holds for any $x$.

In Proposition \ref{complement-case}, we will give basic properties of
functions in $\co\cflmv$ and of the operator $\ominus$.  To describe the proposition, we need to introduce a new function class, denoted by $\nfamv$, which is
defined in a similar way of introducing $\cflmv$ using,   in place of npda's,  {\em one-way (one-head)
nondeterministic finite automata} (or nfa's,  in short) with write-only output
tapes, provided that the termination condition (\ie all
computation paths terminate in linear time) must hold.

\begin{proposition}\label{complement-case}
\begin{enumerate}\vs{-1}
  \setlength{\topsep}{-2mm}%
  \setlength{\itemsep}{0mm}
  \setlength{\parskip}{0cm}%

\item $\co(\co\cflmv) = \cflmv$.

\item $\co\cflmv = \nfamv\ominus \cflmv$.

\item $\cflmv\ominus \cflmv = \cflmv\wedge \co\cflmv$.

\item $\cflmv\neq \co\cflmv$. The same holds for $\cflmvt$.
\end{enumerate}
\end{proposition}

\begin{proofsketch}
We will show only (2). ($\supseteq$)
Let $f\in \nfamv\ominus\cflmv$ and take two functions $h\in\nfamv$ and
$g\in\cflmv$ for which $f(x) = h(x) - g(x)$ (set difference) for all inputs
$x\in\Sigma^*$.
Choose a linear polynomial $p$ satisfying that, for every pair $(x,y)$, $y\in
h(x)\cup g(x)$ implies $|y|\leq p(|x|)$.  By the definition of $f$, it holds
that $f(x) = \Sigma^{\leq p(|x|)} - (g(x)\cup (\Sigma^{\leq p(|x|)} - h(x)))$
for all $x$. For simplicity, we define
$r(x) = g(x)\cup (\Sigma^{\leq p(|x|)} -
h(x))$ for every $x$. It thus holds that $f(x) = \Sigma^{\leq p(|x|)} - r(x)$.
It is not difficult to show that $r$ is in $\cflmv$.

($\subseteq$)
Let $f\in\co\cflmv$. There are a linear polynomial $p$ and a function $g\in
\cflmv$ satisfying that $f(x) = \Sigma^{\leq p(|x|)} - g(x)$ for all $x$. We set
$h(x) = \Sigma^{\leq p(|x|)}$.
Since $h\in\nfamv$, we conclude that $f$ belongs to $\nfamv\ominus \cflmv$.
\end{proofsketch}

Another basic operation used for functions is {\em functional composition}.
The functional composition $f\circ g$ of two functions $f$ and $g$ is defined
as $(f\circ g)(x) = \bigcup_{y\in g(x)}f(y)$ for every input $x$. For two
function classes $\FF$ and $\GG$, let $\FF\circ \GG =\{ f\circ g \mid
f\in\FF, g\in\GG\}$. In particular, we inductively define $\cflsv^{(1)} =
\cflsv$ and $\cflsv^{(k+1)} = \cflsv\circ \cflsv^{(k)}$ for each index
$k\geq1$.  For instance, the function $f(x)=\{xx\}$ defined for any $x\in\Sigma^*$ belongs to
$\cflsv^{(2)}$. This fact yields, \eg
$\cflsvtwo\subseteq \cflsv^{(4)}$.
Unlike NP function classes (such as, NPSV and NPMV), $\cflsvt$ (and therefore
$\cflsv$ and $\cflmv$) is not closed under functional composition.

\begin{proposition}\label{DUP-complexity}
$\cflsvt\neq \cflsvt^{(2)}$. (Also for $\cflsv$ and $\cflmv$.)
\end{proposition}

\begin{proofsketch}
Let $\Sigma=\{0,1,\natural\}$ be our alphabet and define
$f_{dup\natural}(x)=\{x\natural x\}$ for any input $x\in\{0,1\}^*$ and
$f_{dup\natural}(x)=\{\lambda\}$ for any other inputs $x$.
It is not difficult to show that $f_{dup\natural}\in\cflsvt^{(2)}$.
To show that $f_{dup\natural}\not\in\cflsvt$, we assume
that $f_{dup\natural}\in\cflsvt$.
Let $DUP_{\natural}$ be a ``marked'' version of $DUP$ (duplication), defined as
$DUP_{\natural} = \{x\natural x\mid x\in\{0,1\}^*\}$.
It holds that, for every $w\neq\lambda$, $w\in\ran(f_{dup\natural})$ iff there
is a string $x$ such that $w\in f_{dup\natural}(x)$ iff $w\in DUP_{\natural}$.
Thus, $DUP_{\natural}\cup\{\lambda\} = \ran(f_{dup\natural})$.
Note that $DUP_{\natural}\in\cfl$ since $f_{dup\natural}\in\cflsvt$. However,
it is well-known that $DUP_{\natural}\notin\cfl$. This leads to a
contradiction.
Therefore, we conclude that $f_{dup\natural}\not\in \cflsvt$.
\end{proofsketch}

To examine the role of functions in a process of recognizing a given language,
a {\em  $/\!/$-advice operator},
defined by K\"{o}bler and Thierauf \cite{KT94}, is quite useful.
Given a class $\FF$ of functions, a language $L$ is in $\cfl/\!/\FF$ if there
exists a language $B\in\cfl$ and a function $h\in\FF$ satisfying $L=\{x\mid
\exists\,y\in h(x) \text{ s.t. } \track{x}{y}\in B\}$.
Analogously, $\reg/\!/\FF$ is defined using $\reg$ instead of $\cfl$.
This operator naturally extends a $/$-advice operator of \cite{TYL10,Yam08}.

In the polynomial-time setting, it holds that $\np\cap\co\np= \p/\!/\npsvt$
\cite{KT94}. A similar equality, however, does not hold for $\cfl$ functions.

\begin{proposition}\label{//-cflsvt}
\begin{enumerate}\vs{-1}
  \setlength{\topsep}{-2mm}%
  \setlength{\itemsep}{0mm}
  \setlength{\parskip}{0cm}%

\item $\reg/\!/\nfasvt \nsubseteq \cfl$ and
      $\cfl\nsubseteq \reg/\!/\nfamv$.

\item $\reg/\!/\nfasvt$ is closed under complement but $\reg/\!/\nfamv$ is
    not.

\item $\cfl\cap\co\cfl \subsetneqq  \reg/\!/\cflsvt$.
\end{enumerate}
\end{proposition}

\begin{proofsketch}
(1) Note that the language $DUP_{\#}=\{x\# x\mid x\in\{0,1\}^*\}$ (duplication) falls into $\reg/\!/\nfasvt$ by setting $h(x\# y)=y$ and $B=\{\track{x}{x}\mid
x\in\{0,1\}^*\}$. The key idea is the following claim. (*) A language $L$ is in
$\reg/\!/\nfamv$ iff $L$ is recognized by a certain {\em one-way two-head
(non-sensing) nfa} (or an nfa(2), in short) with $\lambda$-moves. See a survey,
\eg \cite{HKM08}, for this model.
Now, consider $L_{pal}=\{x\# x^R\mid x\in\{0,1\}^*\}$ (palindromes). Since
$L_{pal}$ cannot be recognized by any nfa(2), it follows that $L_{pal}\not\in
\reg/\!/\nfamv$.

(2) The non-closure property of $\reg/\!/\nfamv$ follows from a fact that the
class of languages recognized by nfa(2)'s is not closed under complement. Use
the above claim (*) to obtain the desired result.
\end{proofsketch}

Bwfore closing this section, we exhibit a simple structural difference between
languages and functions. It is well-known that all languages over the unary alphabet
$\{1\}$ in $\cfl$ belong to $\reg$. On the contrary, there is a function
$f:\{1\}^*\rightarrow\{0,1\}^*$ such that $f$ is in $\cflmv$ but not in
$\nfamv$.

\vs{-1}
\section{Oracle Computation and Two Relativizations}\label{sec:relativization}

{\em Oracle computation} is a natural extension of ordinary stand-alone computation
by providing external information by way of {\em query-and-answer}  communication.
Such oracle computation can realize various forms of relativizations, including
many-one and Turing relativizations and thus introduce relativized languages
and functions. By analogy with relativized NP functions (\eg \cite{Sel96}),  let us consider many-one and Turing relativizations of CFL functions.
The first notion of {\em many-one relativization} was discussed for languages in automata
theory \cite{Kob69,Yam14a} and we intend to extend it to CFL
functions.  Given any language $A$ over alphabet $\Gamma$, a function
$f:\Sigma^*\rightarrow \PP(\Gamma^*)$ is in $\cflmv_{m}^{A}$ (or
$\cflmv_{m}(A)$) if  there exists an npda $M$ with two  write-only  tapes (one
of which is a standard output tape and the other is a {\em query tape}) such that, for
any $x\in\Sigma^*$,
(i) along any accepting computation path $p$ of $M$ on $x$ (notationally, $p\in
ACC_{M}(x)$), $M$ produces a query string $y_p$ on the query tape as well as  an output string $z_p$ on the output tape and (ii) $f(x)$ equals the set
$\{ z_p \mid y_p\in A, p\in ACC_{M}(x) \}$. Such an $M$ is referred to as an
{\em oracle npda}.
Given any language family $\CC$, we further set $\cflmv_m^{\CC}$ (or
$\cflmv_{m}(\CC)$) to be $\bigcup_{A\in\CC}\cflmv_m^A$. Similarly, we  define
$\cflsv_{m}^{A}$ and $\cflsv_m^{\CC}$ by additionally demanding the size of
each output string set is at most $1$.
Using $\cflsv_{m}^{A}$, a relativized language family $\cfl_{m}^{A}$ defined
in \cite{Yam14a} can be expressed as $\cfl_{m}^{A}=\{L\mid \eta_{L}\in\cflsv_{m}^A\}$.

\begin{lemma}\label{many-one-case}
\begin{enumerate}\vs{-1}
  \setlength{\topsep}{-2mm}%
  \setlength{\itemsep}{0mm}
  \setlength{\parskip}{0cm}%

\item  $\cflmv_{m}^{\reg} = \cflmv$ and $\cflsv_{m}^{\reg} = \cflsv$.

\item $\cflsv^{(k+1)} = \cflsv_{m}^{\cfl(k)}$.

\end{enumerate}
\end{lemma}


\begin{proposition}\label{upper-//-cflsvt}
\begin{enumerate}\vs{-1}
  \setlength{\topsep}{-2mm}%
  \setlength{\itemsep}{0mm}
  \setlength{\parskip}{0cm}%

\item $\reg/\!/\cflsvt \subseteq \cfl_{m}^{\cfl(2)}\cap
    \co\cfl_{m}^{\cfl(2)}$.

\item  $\cfl/\!/\cflsvt \subseteq \cfl_{m}^{\cfl(3)}$.
\end{enumerate}
\end{proposition}

\begin{proofsketch}
We will prove only (2). For convenience, we write $[x,y]^T$ for $\track{x}{y}$ in this proof.
Let $\cfl_{m[1]}^{A} = \cfl_{m}^{A}$ and $\cfl_{m[k+1]}^{A} =
\cfl_{m}(\cfl_{m[k]}^{A})$ for every index $k\geq1$ \cite{Yam14a}.
Let $L\in\cfl/\!/\cflsvt$ and take an npda $M$ and a function $h\in\cflsvt$
for which  $L =\{x\mid \text{ $M$ accepts $[x,h(x)]^T$ }\}$.  Let $M'$ be an
npda with an output tape computing $h$. An oracle npda $N$ is also defined as follows. On input $x$,
$N$ simulates $M'$ on $x$ and nondeterministically produces  $[\tilde{x},\tilde{y}]^T$ on its query
tape when $M'$ outputs $y$, where $\tilde{x}$ and $\tilde{y}$ are appropriate
$\natural$-extensions of $x$ and $y$, respectively. An oracle  $A$
receives  a
$\natural$-extension $[\tilde{x},\tilde{y}]^T$ and decides whether $M$
accepts $[x,y]^T$ by removing all $\natural$s.
Clearly, $L$ belongs to $\cfl_{m}^{A}$ via $N$.
Define another npda $N_1$.  On input $w= [\tilde{x},\tilde{y}]^T$, $N_1$
simulates $M$ on $[x,z]^T$ by guessing $z$ symbol by symbol. At the same time,
it writes $[\tilde{y}',\tilde{z}]^T$ on a query tape
and accepts $w$
exactly when $M$ enters an accepting state. Let $B =\{
[\tilde{y}',\tilde{z}]^T\mid y=z\}$. Note that $B$ is in $\cfl_{m}^{\cfl}$.
Hence, $A$ is in $\cfl_{m}^{B}\subseteq \cfl_{m[2]}^{\cfl}$, and thus $L$ is in
$\cfl_{m[3]}^{\cfl}$.  Since $\cfl_{m[3]}^{\cfl} = \cfl_{m}^{\cfl(3)}$
\cite{Yam14a}, the desired conclusion follows.
\end{proofsketch}


The second relativization is {\em Turing relativization}. A multi-valued
partial function $f$ belongs to $\cflmv_{T}^{A}$ (or $\cflmv_{T}(A)$) if there
exists an oracle npda $M$ having {\em three extra inner states}
$\{q_{query},q_{yes},q_{no}\}$ that satisfies the following three conditions:
on each input $x$, (i) if $M$ enters a query state $q_{query}$, then a valid
string, say, $s$ written on the query tape is sent to $A$ and, automatically,
{\em the content of the query tape becomes blank} and {\em the tape head
returns to the start cell}, (ii) oracle $A$ sets $M$'s inner state to $q_{yes}$
if $s\in A$ and $q_{no}$ otherwise, and (iii) all computation paths of $M$
terminate in time $O(n)$ {\em no matter what oracle is used}. Obviously,
$\cflmv_{T}^{A}= \cflmv_{T}^{\overline{A}}$ holds for any oracle $A$.
Define $\cflmv_{T}^{\CC}$ (or $\cflmv_{T}(\CC)$) to be  the union
$\bigcup_{A\in\CC}\cflmv_{T}^A$ for a given language family $\CC$.

Analogously to the well-known {\em NPMV hierarchy}, composed of
$\Sigma^{\mathrm{p}}_{k}\mathrm{MV}$ and $\Pi^{\mathrm{p}}_{k}\mathrm{MV}$ for
$k\in\nat^{+}$ \cite{Sel96}, we inductively
define $\sigmacflmv{1} =\cflmv$, $\picflmv{k} = \co\sigmacflmv{k}$, and
$\sigmacflmv{k+1} = \cflmv_{T}({\picfl{k}})$ for every index $k\geq1$.
In a similar fashion, we define $\sigmacflsv{k}$ using $\cflsv_{T}^{A}$ in place of
$\cflmv_{T}^{A}$.
The above {\em CFLMV hierarchy} is useful to scaling the computational
complexity of given functions. For example, the function
$f(w)=\{x\in\{0,1\}^*\mid \exists u,v\,[w=uxxv]\}$ for every $w\in\{0,1\}^*$
belongs to $\sigmacflmv{2}$.  Moreover, it is possible to show that  $\cflmv\cup \co\cflmv
\subseteq \cflmv\ominus \cflmv \subseteq \sigmacflmv{4}$.

\begin{proposition}\label{lin-space-bound}
Each function in $\bigcup_{k\in\nat^{+}}\sigmacflsvt{k}$ can be computed by an
appropriate $O(n)$ space-bounded multi-tape deterministic Turing machine.
\end{proposition}

\begin{proofsketch}
It is known in \cite{Yam14a} that $\sigmacfl{k}\subseteq\mathrm{DSPACE}(O(n))$
for every $k\geq1$. It therefore suffices to show that, for any fixed language
$A\in \mathrm{DSPACE}(O(n))$, every function $f$ in $\cflsvt^{\!A}$ can be
computed using $O(n)$ space. This is done by a direct simulation of $f$ on a
multi-tape Turing machine.
\end{proofsketch}

For $k\geq3$, it is possible to give the exact characterization of
$\reg/\!/\sigmacflsvt{k}$. This makes a sharp contrast with Proposition
\ref{//-cflsvt}(3).

\begin{proposition}\label{k-level-//-sv}
For every index  $k\geq3$,
$\sigmacfl{k}\cap \picfl{k} = \reg/\!/\sigmacflsvt{k}$.
\end{proposition}

\begin{proofsketch}
By extending the proof of Proposition \ref{//-cflsvt}(3), it is possible to
show that $\sigmacfl{k}\cap \picfl{k} \subseteq \reg/\!/\sigmacflsvt{k}$.
Similarly to Proposition \ref{//-cflsvt}(2), it holds that
$\reg/\!/\sigmacflsvt{k}$ is closed under complement.
It thus suffices to show that $\reg/\!/\sigmacflsvt{k} \subseteq \sigmacfl{k}$.
This can be done by a direct simulation.
\end{proofsketch}

\begin{lemma}\label{relation-CFL-CFLSV}
Let $e\geq2$ and $k\geq1$.
$\sigmacflsv{k} = \sigmacflsv{k+1}$ $\Rightarrow$
$\sigmacfl{k}=\sigmacfl{k+1}$ $\Rightarrow$
$\sigmacflsv{k+1}=\sigmacflsv{k+e}$.
\end{lemma}

\begin{proofsketch}
Assuming $\sigmacflsv{k}=\sigmacflsv{k+1}$, let us take any language
$A\in\sigmacfl{k+1}$ and consider $\eta_{A}$. It
is not difficult to show that, for any index $d\in\nat^{+}$,
$A\in\sigmacfl{d}$ iff  $\eta_{A}\in\sigmacflsv{d}$. Thus,
$\eta_{A}\in\sigmacflsv{k+1} = \sigmacflsv{k}$. This implies that
$A\in\sigmacfl{k}$.
Next, assume that $\sigmacfl{k}=\sigmacfl{k+1}$. It is proven in \cite{Yam14a}
that $\sigmacfl{k}=\sigmacfl{k+1}$ iff $\sigmacfl{k}=\sigmacfl{k+e}$ for all
$e\geq1$.  Hence, for every $e\geq2$, we obtain $\sigmacfl{k}=\sigmacfl{k+e-1}$, which is
equivalent to $\picfl{k}=\picfl{k+e-1}$. It then follows that $\sigmacflsv{k+e}
= \cflsv_{T}({\picfl{k+e-1}}) = \cflsv_{T}({\picfl{k}}) = \sigmacflsv{k+1}$.
\end{proofsketch}

Regarding refinement, from the proof of \cite[Theorem 3]{Kob69} follows
$\nfamv\sqsubseteq_{ref}\nfasv$. This result leads to $\nfamv\circ \cflsv
\sqsubseteq_{ref}\cflsv$ in \cite{KSY07}.
By a direct simulation, nevertheless, it is possible to show that
$\sigmacflmv{k}\sqsubseteq_{ref} \sigmacflsv{k+1}$ for every $k\geq1$.
Lemma \ref{relation-CFL-CFLSV} together with this fact leads to the following
consequence.

\begin{proposition}\label{CFLMV-included-CFLSV}
Let $k\geq2$. If $\sigmacfl{k}=\sigmacfl{k+1}$, then
$\sigmacflmv{k+1}\sqsubseteq_{ref} \sigmacflsv{k+1}$.
\end{proposition}

Recently, it was shown in \cite{Yam14b} that
$\cflmv\not\sqsubseteq_{ref}\cflsv$ holds. However, it is not known if this can
be extended to every level of the CFLMV hierarchy.

\section{Optimization Functions}\label{sec:optimization}

An {\em optimization problem} is to find an optimal feasible solution that
satisfy a given condition. Krentel \cite{Kre88} studied the complexity of those
optimization problems. Analogously to $\optp$ of Krentel, we define $\optcfl$
as a collection of {\em single-valued total functions}  $f:\Sigma^*\rightarrow
\Gamma^*$ such that there exists an npda $M$ and an
$opt\in\{\text{maximum,minimum}\}$ for which, for every string $x\in\Sigma^*$,
$f(x)$ denotes the $opt$ output string of $M$ on input $x$ along an appropriate
accepting computation path, assuming that $M$ must have at least one accepting
computation path.
Here, we use the {\em dictionary (or alphabetical) order} $<$ over $\Gamma^*$
(e.g., $abbe<abc$ and $ab<aba$) instead of the lexicographic order to
compensate the npda's limited ability of comparing two strings {\em from left to
right}. For example, the function $f(w) =\max\{PAL_{sub}(w)\}$ for
$w\in\{0,1\}^*$, where $PAL_{sub}$ is defined in Section \ref{sec:introduction}, is a member of $\optcfl$. It holds that
$\cflmvt\sqsubseteq_{ref}\optcfl$.

\begin{proposition}\label{OptCFL-bounds}
$\cflsvt\subsetneqq \optcfl\subseteq \sigmacflsvt{4}$.
\end{proposition}

The first part of Proposition \ref{OptCFL-bounds} comes from a fact that the
function  $f(w)=\max\{g(w)\}$, where $g(w)=\{\lambda\}\cup \{x_iy_i\mid
w=x_1\natural x_2\natural x_3\# y_1\natural y_2\natural y_3, x_i=y_i^{R},
i\in\{1,2,3\}\}$, is in $\optcfl$ but not in $\cflsv_t$. This latter part is
proven by applying the functional pumping lemma.

Note that, in the polynomial-time setting, a much sharper upper-bound of
$\optp\subseteq\Sigma^{\mathrm{p}}_{2}\mathrm{SV_t}$ is known.
Similarly to $\optcfl$, let us define $\optnfa_{EL}$ using nfa's $M$ instead of
npda's with an extra condition that $M(x)$ outputs only strings of the {\em
equal length}. This new class is located within $\sigmacflsvt{2}$.

\begin{proposition}\label{opt-NFA-bound}
$\optnfa_{EL}\subseteq \sigmacflsvt{2}$.
\end{proposition}

\begin{proofsketch}
Let $f\in \optnfa_{EL}$ and take an underlying nfa $N$ that forces $f(x)$ to  equal
$\max\{N(x)\}$ for every $x$. Define an oracle npda $M_1$ to simulate $N$ on
$x$ and output, say, $y$. Simultaneously, query $y\#x^R$ using a stack wisely.
If its oracle answer is $1$, enter an accepting state; otherwise, reject.
Make another npda $M_2$ receive $y\#x^R$, simulate $N^R$ on $x^R$, and compare  its
outcome with $y^R$, where the notation $N^R$ refers to an nfa that reverses the
computation of $N$.
\end{proofsketch}

\begin{proposition}\label{cfl/opt-complexity}
\begin{enumerate}\vs{-1}
  \setlength{\topsep}{-2mm}%
  \setlength{\itemsep}{0mm}
  \setlength{\parskip}{0cm}%

\item $\cfl\cup\co\cfl\subseteq \reg/\!/\optcfl \subseteq
    \sigmacfl{4}\cap\picfl{4}$.

\item  $\cfl/\!/\optcfl \subseteq \sigmacfl{5}$.
\end{enumerate}
\end{proposition}

\begin{proofsketch}
We will show only (1). Note that $\reg/\!/\optcfl$ is
closed under complementation. Let $L\in\cfl$ and take an npda $M$ recognizing
$L$. Define $N_1$ as follows. On input $x$, guess a bit $b$. If $b=0$, then output $0$
in an accepting state. Otherwise, simulate $M$ on $x$ and output $1$ along only
accepting computation paths of $M$. Let $h(x)$ be $\max\{N_1(x)\}$ for all $x$'s. It follows that $L=\{\track{x}{h(x)}\mid h(x)\neq\setempty\}$.
This proves that $L\in \reg/\!/\optcfl$. Next, let $L\in\reg/\!/\optcfl$. Since
$\optcfl\subseteq \sigmacflsvt{4}$, we obtain $L\in\reg/\!/\sigmacflsvt{4}$. By
Proposition \ref{k-level-//-sv}, this implies that  $L$ belongs to
$\sigmacfl{4}\cap\picfl{4}$.
\end{proofsketch}

\let\oldbibliography\thebibliography
\renewcommand{\thebibliography}[1]{%
  \oldbibliography{#1}%
  \setlength{\itemsep}{0pt}%
}
\bibliographystyle{plain}

\end{document}